\documentclass[prl,twocolumn,showpacs,superscriptaddress]{revtex4-1}
\usepackage{graphicx,amssymb,amsmath,psfrag,color}
\begin{document}
\definecolor{darkgreen}{rgb}{0,0.5,0}
\newcommand{\be}{\begin{equation}}
\newcommand{\ee}{\end{equation}}
\newcommand{\jav}[1]{\textcolor{red}{#1}}

\title{Unusual spin dynamics in topological insulators}

\author{Bal\'azs D\'ora}
\email{dora@eik.bme.hu}
\affiliation{BME-MTA Exotic  Quantum  Phases Research Group, Budapest University of Technology and
  Economics, 1521 Budapest, Hungary}
\affiliation{Department of Physics, Budapest University of Technology and  Economics, 1521 Budapest, Hungary}
\author{Ferenc Simon}
\affiliation{Department of Physics, Budapest University of Technology and  Economics, 1521 Budapest, Hungary}

\date{\today}

\begin{abstract}
The dynamic spin susceptibility (DSS) has a ubiquitous Lorentzian form in conventional materials with weak spin orbit coupling,
whose spectral width characterizes the spin relaxation rate. We show that DSS has an unusual non-Lorentzian form in
topological insulators, which are characterized by strong SOC.
At zero temperature, the high frequency part of DSS is universal and increases in certain directions as $\omega^{d-1}$ with $d=2$ and
3 for surface states and Weyl semimetals, respectively, while for helical edge states, the interactions renormalize the exponent as
$d=2K-1$ with $K$ the Luttinger-liquid parameter. As a result, spin relaxation rate cannot be deduced from the DSS in contrast to the
case of usual metals, which follows from the strongly entangled spin and charge degrees of freedom in these systems. These parallel with the
 optical conductivity of neutral graphene.

\end{abstract}


\maketitle

\emph{Introduction.} Strong correlation effects manifest as unusual behavior of physical
response functions. Of these, the frequency and momentum dependent spin susceptibility,
$\chi(q,\omega)$, played a pivotal role in the study of e.g.
high-temperature superconductors \cite{neutron}, spin-ice compounds \cite{spinice}, and the
fundamental description of magnetic resonance experiments in correlated
systems \cite{AlloulPRL1989}. This response function is available
experimentally using \textit{ac} magnetization measurements, neutron scattering,
magnetic resonance, M\"ossbauer spectroscopy, spin-resolved STM, or microwave cavity
perturbation experiments. Common to these method is that it is difficult to deduce the full $\omega$
dependent signal,  the analysis of experiment therefore relies on the
theoretical description of the susceptibility.

The long wavelength spin susceptibility, $\chi(q\rightarrow 0,\omega)$, called the \textit{ac} or dynamic spin susceptibility (DSS), indicates dissipative processes and
 remains in the focus of interest when studying the nature of correlations in emergent
materials, such as e.g. those manifesting the spin-liquid phase \cite{SpinLiquidMoessner}.
DSS is also important in identifying the transition temperature of spin-glasses\cite{mulder} and superconductors\cite{konigschindler}, characterizing superparamagnetism of small ferromagnetic nanoparticles\cite{superpara}, or
examining the nature of magnetic phase transitions.
Another highly relevant reason to study DSS is that it provides a measure of spin-relaxation
rate, whose knowledge is in turn important for spintronics applications \cite{FabianRMP}.
DSS is characterized in the usual materials (where spin-orbit interaction is small) by a
Lorentzian, which is peaked at the Zeeman energy and whose linewidth provides a direct measure of the spin-relaxation rate.

A common feature of the previous cases is the weak spin orbit coupling (SOC). However, SOC is usually the dominant energy scale in topological insulators\cite{hasankane,rmpzhang}
which strongly entangles their magnetic properties with their charge response. As a result, unusual,
non-Lorentzian behavior of the DSS might occur. Here, we study DSS in topological insulators in
the full temperature, doping, magnetic field and frequency range. We do find a non-Lorentzian
form of the DSS and most surprisingly a non-zero value of the DSS even in the large frequency limit. This
implies that the spin-relaxation rate cannot be determined from the DSS, much as its knowledge is
desired for prospective spintronics applications. This result is understood in analogy to the case
of optical conductivity of \emph{neutral} graphene: it does not follow the usual Drude-Lorentz form due to two-band
 excitations, therefore it cannot be used to determine the momentum relaxation rate \cite{opticalgeim}.
We argue that the non-Lorentzian form of response functions is a new hallmark of topological insulators.

In the following, we study four known realizations of topological insulators: i) the spin Hall edge state, ii) its interacting counterpart, the helical liquid in 1D, iii) 2D helical Dirac fermions, iv) and the Weyl semimetal in 3D.

\emph{1D Dirac Hamiltonian: the spin-Hall edge state.} We consider the spin-filtered edge states of a quantum spin-Hall insulator\cite{kanemele1,bernevig,konig}, whose effective Hamiltonian is
\begin{gather}
H_{1d}=vS_zp+\Delta S_x
\label{hamilton1d}
\end{gather}
with energy spectrum $E_\pm(p)=\pm\sqrt{(vp)^2+\Delta^2}$ and $\Delta$ is the Zeeman term from a static magnetic field.

The DSS requires the calculation of the spin response function, which reads in the time domain as
\begin{gather}
\chi_{ab}(t)=i\Theta(t)\langle S_a(t)S_b(0)-S_b(0)S_a(t)\rangle,
\end{gather}
where $a,b=x$, $y$, or $z$ 
and $S_a(t)=\exp(-iH_{1d}t)S_a\exp(iH_{1d}t)$ can be calculated using the matrix structure of $H_{1d}$, similarly to Ref. \cite{dorarabi}.
A given momentum plays the role of an effective magnetic field, which acts on the physical spin.
Therefore, the knowledge of the $\chi_{ab}(t)$ correlator yields directly the DSS.
Using the eigenfunctions of Eq. \eqref{hamilton1d}, the time dependent correlation function for a given momentum $p$
is calculated, yielding the imaginary part of the DSS at half filling and $T=0$ after Fourier transformation as
\begin{subequations}
\begin{gather}
\chi_{zz}^{\prime\prime}(\omega)= \frac{\Delta^2}{v\omega\sqrt{\omega^2-4\Delta^2}}\Theta(\omega^2-4\Delta^2),\\
\chi_{yy}^{\prime\prime}(\omega)= \frac{\omega}{4v\sqrt{\omega^2-4\Delta^2}}\Theta(\omega^2-4\Delta^2),\\
\chi_{xx}^{\prime\prime}(\omega)= \frac{\sqrt{\omega^2-4\Delta^2}}{4v\omega}\Theta(\omega^2-4\Delta^2),\\
\chi_{yz}^{\prime\prime}(\omega)= \frac{1}{v\pi}F\left(\frac{\omega}{2|\Delta|}\right),
\end{gather}
\end{subequations}
where $F(x)=[(\textmd{arctan}(\sqrt{1-x^2}/x)-\pi/2)\Theta(1-|x|)+\textmd{atanh}(\sqrt{x^2-1}/x)\Theta(|x|-1)]/\sqrt{|x^2-1|}$.
The DSS is thus strongly anisotropic, contains an off-diagonal term and deviates from the ideal Lorentzian form. Depending on the geometry,
the DSS diverges or vanishes at the gap edge and approaches a finite constant value or vanishes with increasing frequency.

In the $\Delta=0$ limit, $S_z$ is conserved ($[S_z,H_{1d}]=0$), therefore $\chi_{zz}^{\prime\prime}(\omega)=\chi_{yz}^{\prime\prime}(\omega)=0$, while
$\chi_{xx}^{\prime\prime}(\omega)=\chi_{yy}^{\prime\prime}(\omega)=\textmd{sgn}(\omega)/4v$, which is the typical density of states in 1D.

The electric current operator is given by $j_x=evS_z$, therefore
 the optical conductivity of the spin-Hall edge state measure directly $\chi_{zz}^{\prime\prime}(\omega)/\omega$.
Note that the other components of DSS are not accessible by optical means.
Additionally, a finite \textit{ac} \emph{electric} current can be induced along the edge in the presence of an \textit{ac} \emph{magnetic}
field in the $y$ direction due to the finite value of $\chi_{yz}(\omega)$, as
a manifestation of the magnetoelectric effect\cite{hasankane,rmpzhang}. In particular, Re$\chi_{yz}(\omega\gtrsim 2\Delta)\sim \sqrt{\Delta/(\omega-2\Delta)}$.

The results for the diagonal susceptibilities can be extended to finite doping and
 temperature by multiplying the calculated $\chi_{aa}$'s by $\textmd{sgn}(\omega)(f(\mu-\frac \omega 2)-f(\mu+\frac \omega 2))$ (except for the case of the helical liquid),
 where $f(E)=1/(\exp(E/k_BT)+1)$ with $T$ the temperature and $\mu$ the chemical potential.
At $T=0$, a finite chemical potential introduces an additional gap of $2|\mu|$, and leaves the rest intact. At high
temperature, it gives a $|\omega|/4T$ multiplicative factor to the susceptibilities.

A finite perpendicular magnetic field $\Delta$ opens up a gap in the spectrum, and the resulting state becomes immune with respect to interactions as long as $|\mu|\ll\Delta$.
In the absence of the gap, the density of states is finite for arbitrary chemical potential, and the interactions profoundly alter the low energy excitations, as is customary in 1D\cite{giamarchi}.
The results obtained below apply also in the case of a finite gap, unless $\mu\sim\Delta\sim\omega$.

\emph{Helical liquid.} The helical edge state of the spin-Hall insulator forms a helical liquid, when electron-electron interaction is taken
into account, resembling to a spinless Luttinger liquid (LL)\cite{cwu,cxu,hohenadler}.
The Hamiltonian in Eq. \eqref{hamilton1d} is rewritten in second quantized form as \cite{hou}
\begin{gather}
H_0=iv\int dx\left(R^+_\uparrow(x)\partial_xR_\uparrow(x)-L^+_\downarrow(x)\partial_xL_\downarrow(x)\right),
\end{gather}
which is a peculiar half of a spinful LL, lacking the $R_{\downarrow}$ and $L_\uparrow$ operators.

The time reversal invariant electron-electron interaction consists of the chiral ($g_4$)  and the forward scattering ($g_2$) terms,
\begin{gather}
H_{int}=\sum_{s=\uparrow,\downarrow}\frac{g_4}{2}\int dx (n_s(x))^2+g_2\int dx ~n_\uparrow(x)n_\downarrow(x)
\end{gather}
with $n_\uparrow(x)=R^+_\uparrow(x)R_\uparrow(x)$ and $n_\downarrow(x)=L^+_\downarrow(x)L_\downarrow(x)$.
These interactions give rise to Luttinger liquid behaviour\cite{cwu,cxu,hohenadler} with LL parameter $K$ and
 renormalized velocity $v_F$, and $K=1$ and $v_F=v$ in the non-interacting limit.
The bosonized Hamiltonian reads as
\begin{gather}
H=\frac{v_F}{4\pi}\int dx \left[\frac 1K \left(\partial_x\varphi\right)^2+K\left(\partial_x\theta\right)^2\right],
\end{gather}
with the dual fields $\theta$ and $\varphi$, satisfying $[\varphi(x),\theta(y)]=i\frac \pi 2 \textmd{sgn}(y-x)$.

\begin{figure}[h!]
\centering
\psfrag{p}[b][t][1][0]{$p$}
\psfrag{E}[lb][][1][0]{$E(p)$ }
\psfrag{mu}[b][t][1][0]{$\mu$}
\psfrag{t1}[][][1][0]{\color{darkgreen}$q=2k_F$, $\omega=0$}
\psfrag{t2}[][][1][90]{\color{magenta}$q=0$, $\omega=2\mu$}
\psfrag{up}[][][1.1][0]{\color{blue}$\uparrow$}
\psfrag{down}[][][1.1][0]{\color{red}$\downarrow$}
\includegraphics[width=4.6cm]{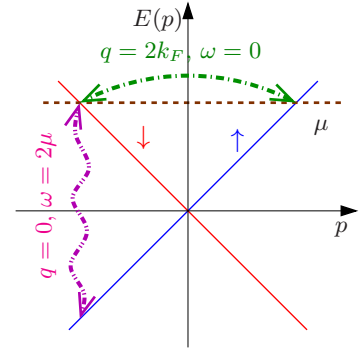}
\caption{(Color online)
The two possible spin-flip processes in the helical liquid, the blue (up spin) and red (down spin) lines denote the bare, spin filtered dispersion.
The $q=0$ process,  corresponding the vertical magenta line, is absent in a normal LL and requires
 a finite frequency threshold $2\mu$,
while the green arrow denotes a gapless, $q=2k_F$ momentum transfer process, which does not contribute to DSS, except for $\mu=0$, when these two processes coincide.}
\label{spechel}
\end{figure}

\begin{figure}[h!]
\centering
\psfrag{x}[t][][1][0]{$\omega/2\mu$}
\psfrag{y}[b][t][1][0]{$\chi_{xx}^{\prime\prime}(\omega)4v_F (v_F/2\pi\alpha\mu)^{2K-2}$}
\psfrag{K=1.3}[l][][1][0]{$K=1.3$}
\psfrag{K=1}[l][][1][0]{\color{blue}$K=1 $ }
\psfrag{K=0.7}[l][][1][0]{\color{red}$K=0.7$}
\includegraphics[width=6.6cm]{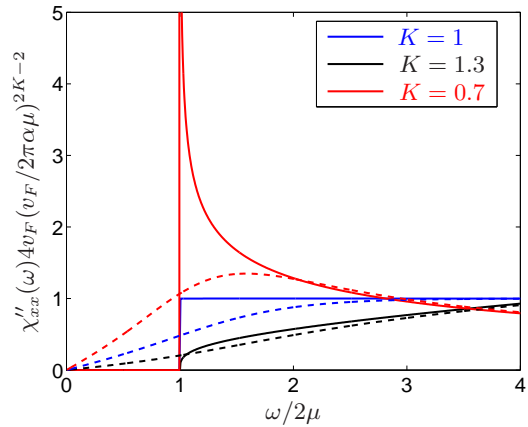}


\caption{(Color online) The dynamical spin susceptibility of the helical liquid is shown for $T=0$ (solid lines) and $T=\mu/2$ (dashed lines) for several values
of the LL parameter.
}
\label{esrhl}
\end{figure}

The DSS of the helical liquid is evaluated similarly to the $2k_F$ \emph{charge} susceptibility of a spinless LL\cite{giamarchi}.
The spin flip operator is translated to the bosonic language as $R^+_\uparrow(x)L_\downarrow(x)\sim \exp(-2ik_Fx+2i\varphi(x))$.
In the absence of perpendicular magnetic field, we obtain
$\chi_{zz}^{\prime\prime}(\omega)=0$ and $\chi_{xx}^{\prime\prime}(\omega)=\chi_{yy}^{\prime\prime}(\omega)$ as
\begin{gather}
\chi_{xx}^{\prime\prime}(\omega)= \frac{\sin(\pi K)}{4v_F\pi^2}\left(\frac{2\pi\alpha T}{v_F}\right)^{2K-2}\times\nonumber\\
\times\textmd{Im}\left[\prod_{r=\pm}B\left(\frac{\omega-2r\mu}{4i\pi T}+\frac{K}{2},1-K\right)\right],
\label{spinhelical}
\end{gather}
where $B(x,y)=\Gamma(x)\Gamma(y)/\Gamma(x+y)$ is the Euler integral of the first kind with $\Gamma(x)$ being the Euler's integral of the second kind\cite{gradstein},
$\alpha$ is a short distance regulator and $v_F/\alpha$ represent a high energy cutoff and is shown in Fig. \ref{esrhl} for some representative cases.
At $T=0$, Eq. \eqref{spinhelical}  exhibits the typical power law correlation function of a LL as
\begin{gather}
\chi_{xx}^{\prime\prime}(\omega,T=0)=\frac{\textmd{sgn}(\omega)}{4v_F\Gamma^2(K)}\left(\frac{\alpha}{2v_F}\right)^{2K-2}\times\nonumber\\
\times\left(\omega^2-4\mu^2\right)^{K-1}\Theta(\omega^2-4\mu^2),
\end{gather}
while in the high temperature limit with $T\gg \omega,\mu$, it yields
\begin{gather}
\chi_{xx}^{\prime\prime}(\omega)= \frac{\omega\alpha}{v_F^2\pi}\left(\frac{2\pi\alpha T}{v_F}\right)^{2K-3}\frac{\Gamma^4(K/2)}{\Gamma^2(K)}.
\end{gather}

In spite of the formal similarity to the $2k_F$, finite frequency response of normal LLs, Eq. \eqref{spinhelical} describes a completely different physical process, which
usually involves high energy transfer and is beyond the realm of the LL paradigm.
While the former is gapless in $\omega$ and accounts for a "horizontal" interband process with $2k_F$ momentum transfer,
the latter stems from a $q=0$ "vertical" interband transition and is gapped at $T=0$ with the threshold frequency of interband transition $2\mu$, as shown
in Fig. \ref{spechel}.
Only at $\mu=0$, these two processes become identical.
By the replacement  $2\mu\longrightarrow 2\mu\pm vq$ in Eq. \eqref{spinhelical}, the
full wavevector dependence of the dynamical susceptibility is obtained.
This indicates that the  chemical potential dependent DSS is equivalent to measure the full wavevector dependent
susceptibility, accessible by e.g. neutron scattering.
Similar results were obtained for the dynamical density response function as well\cite{Gangadharaiah}.

This very broad spin response is reminiscent of that in the XXZ Heisenberg model\cite{giamarchi},
which describes frozen charge degrees of freedom due to the strong on site repulsion between
electrons. The helical liquid, on the other hand, operates in the opposite, weakly interacting itinerant electron limit,
but the strong SOC entangles the spin excitations with the charge degrees of freedom, resulting in
 a broad signal.

In particular, a strongly repulsive helical liquid with $K\ll 1$ produces significantly larger spin responses as opposed to its weakly or attractively interacting counterpart:
 the $({2\pi\alpha T}/{v_F})^{2K-2}$ factor significantly enhances/suppresses the spin susceptibility in the repulsive ($K<1$)/attractive ($K>1$) case.
For $K=1$, our previous expressions for the non-interacting case are recovered.

Eq. \eqref{spinhelical}  is to be contrasted to the spin response of a spinful LL, which in the presence of SU(2) invariant
interactions, reduces to $\omega\delta(\omega\pm B)$ with $B$ the Zeeman field, in spite of the
fractionalization of the original fermionic excitations into new type of collective bosonic modes. Departures from this
highly idealized limit imply the inclusion of various SOC terms into the LL Hamiltonian\cite{martino,doraesr} as a weak perturbation on the band structure. Our starting
point, on the other hand, is the completely opposite situation, when the SOC determines and dominates the band structure, therefore the SU(2) spin rotational symmetry is severely broken and
cannot be considered as a weak perturbation.

\emph{2D Dirac Hamiltonian.}
By increasing the dimensionality, the surface states of 3D topological insulators is described by the familiar Dirac equation\cite{hasankane}, given by
\begin{gather}
H_{2d}=v\left(S_xp_y-S_yp_x\right)+\Delta S_z,
\label{hamilton2d}
\end{gather}
where $\Delta$ is a mass gap, stemming from a thin ferromagnetic film covering the
surface of TI or by a perpendicular magnetic field.
The eigenenergies are $E_\pm({\bf p})=\pm\sqrt{(vp)^2+\Delta^2}$.

The time dependent correlation function is obtained similarly to the 1D case, and
the DSS at $T=0$ and half filling is
\begin{subequations}
\begin{gather}
\chi_{xx}^{\prime\prime}(\omega)= \frac{\omega}{16v^2}\left(1+\frac{4\Delta^2}{\omega^2}\right)\Theta(\omega^2-4\Delta^2),\\
\chi_{zz}^{\prime\prime}(\omega)= \frac{\omega}{8v^2}\left(1-\frac{4\Delta^2}{\omega^2}\right)\Theta(\omega^2-4\Delta^2),\\
\chi_{xy}^{\prime\prime}(\omega)=- \frac{\Delta}{4\pi v^2}\ln\left|\frac{\omega+2|\Delta|}{\omega-2|\Delta|}\right|,
\end{gather}
\label{susc2d}
\end{subequations}
and $\chi_{yy}^{\prime\prime}(\omega)=\chi_{xx}^{\prime\prime}(\omega)$.
Note that $\chi_{xy}^{\prime\prime}(\omega)$ is responsible to the "half quantum
Hall effect", i.e. the $e^2/2h$ Hall conductivity in topological insulators\cite{hasankane}.
Since the electric current operator is related to the spin due to the strong SOC, the in-plane optical conductivity satisfies
$\sigma(\omega)\sim \chi_{xx}^{\prime\prime}(\omega)/\omega$, and this also agrees with the interband contribution to the optical conductivity of (gapped) monolayer graphene\cite{sharapov6}.
While $\chi_{xx,xy}^{\prime\prime}(\omega)$ is measurable by optical means as well, the $zz$ component can only be probed
by magnetic susceptibility measurements.
In the $\Delta=0$ limit, the relation $\chi_{zz}(\omega)=2\chi_{xx}(\omega)= \frac{\omega}{8v^2}$ holds where the last expression is the typical density of states of e.g. graphene\cite{castro07}.
The factor 2 follows from the spin structure of Eq. \eqref{hamilton2d}:
$S_z$ sees two perpendicular spin components ($x$ and $y$), which contribute to the response, while an in plane component feels only the other in-plane component but not $S_z$.
Qualitatively similar susceptibilities were derived in Ref. \cite{haolei}.

The effect of a short range electron-electron interaction (e.g. Hubbard model) is practically negligible here, as it
is termed irrelevant in the renormalization group sense and can only renormalize the band parameters in the weak coupling limit.

\begin{figure}[h!]
\centering
\psfrag{x}[t][][1][0]{$\omega/2\Delta$}
\psfrag{y}[b][t][1][0]{$\chi^{\prime\prime}(\omega)8v^2/\Delta$}
\psfrag{t1}[][r][1][0]{$\chi_{zz}^{\prime\prime}$, $T=0$}
\psfrag{t2}[][r][1][0]{$\chi_{xx}^{\prime\prime}$, $T=0$}
\psfrag{t3}[][r][1][0]{$\chi_{xy}^{\prime\prime}$, $T=0$}
\psfrag{t4}[][r][1][0]{$\chi_{zz}^{\prime\prime}$, $T=5\Delta$}
\psfrag{t5}[][r][1][0]{$\chi_{xx}^{\prime\prime}$, $T=5\Delta$}
\psfrag{t6}[][r][1][0]{$\chi_{xy}^{\prime\prime}$, $T=5\Delta$}
\includegraphics[width=7cm]{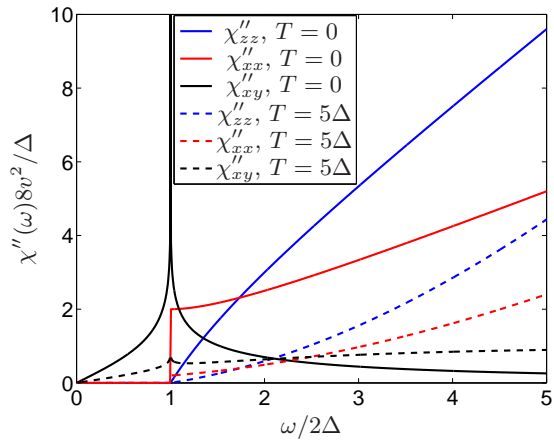}
\caption{(Color online) The dynamical spin susceptibility of the 2D
topological surface state is shown for $T=0$ (solid lines) and $T=5\Delta$ (dashed lines) at half filling.}
\label{esr2d}
\end{figure}

\emph{Weyl semimetal.}
Inspired by the exciting  physics of graphene and topological insulators, nodal semimetals in 3D are currently under investigation\cite{balents, burkov, delplace}.
The Weyl Hamiltonian exhausts all three spin operators as
\begin{gather}
H_W=v\left(S_xp_x+S_yp_y+S_zp_z\right).
\end{gather}
The Zeeman energy simply shifts the position of the zero energy state in the momentum space and does not open a gap in the above Hamiltonian.
The DSS  follows from Eqs. \eqref{susc2d}, after replacing $\Delta$ with $k_z$ and performing the $k_z$ integral, becoming  isotropic and  diagonal as
\begin{gather}
\chi^{\prime\prime}(\omega)= \frac{\omega^2}{24\pi v^3},
\end{gather}
being proportional to the density of states of Weyl semimetals.
Similarly to the previous cases, the  optical conductivity follows as $\sigma(\omega)\sim \chi^{\prime\prime}(\omega)/\omega\sim \omega$ as in Ref. \cite{hosur}.

\emph{Detection.} Experimentally, the DSS is directly measured by the electron spin resonance (ESR) method, whose signal intensity is\cite{Slichterbook}:
\begin{equation}
I_{aa}(\omega)=\frac{B_{\perp}^2\omega}{2 \mu_0}\chi^{\prime\prime}_{aa}(\omega)V,
\end{equation}
where $\mu_0$ is the permeability of the vacuum, $V$ is the sample volume. 
Usually, the conventional ESR method together with the nuclear magnetic resonance (NMR) in solid state  systems has limited importance in 2D and especially 1D
 due to the small  number of available
states (small density of states compared to 3D), which results in  weak signals.
Nevertheless, by considering an ensemble of 1D nanowires and crystals,
the ESR signal can possibly be detected similarly to the NMR spectra\cite{nisson}
of related materials.
Additionally, one can also use the recently proposed source-probe setup to measure the DSS\cite{stano}.
The DSS is accessible in a cold atomic realization of these states (see e.g. Ref. \cite{goldman}), featuring also the tunability of the interaction strength by standard techniques\cite{cazalillarmp},
by measuring the spin-sensitive Bragg signal, yielding the spin-structure factor.

\begin{acknowledgments}

This research has been  supported by the Hungarian Scientific Research Funds Nos. K101244, K105149, K108676, by the ERC Grant Nr. ERC-259374-Sylo and by the Bolyai Program of the HAS.
\end{acknowledgments}

\bibliographystyle{apsrev}
\bibliography{refgraph}

\end{document}